\def\year{2022}\relax
\documentclass[letterpaper]{article} 
\usepackage{aaai22}  
\usepackage{times}  
\usepackage{helvet}  
\usepackage{courier}  
\usepackage[hyphens]{url}  
\usepackage{graphicx} 
\usepackage{natbib}  
\usepackage{caption} 
\DeclareCaptionStyle{ruled}{labelfont=normalfont,labelsep=colon,strut=off} 
\usepackage{algorithm}
\usepackage{algorithmic}

\usepackage{newfloat}
\usepackage{listings}
\floatstyle{ruled}
\newfloat{listing}{tb}{lst}{}
\floatname{listing}{Listing}

\usepackage{comment}
\usepackage{xcolor}

\title{MIRAGE: Multi-model Interface for Reviewing and Auditing \\Generative Text-to-Image AI}
\author{
    Matheus Kunzler Maldaner\textsuperscript{\rm 1}\thanks{This work was conducted while the author was an intern at Carnegie Mellon University.},
    Wesley Hanwen Deng\textsuperscript{\rm 2},\\
    Jason Hong\equalcontrib\textsuperscript{\rm 2},
    Ken Holstein\equalcontrib\textsuperscript{\rm 2},
    Motahhare Eslami\equalcontrib \textsuperscript{\rm 2}
}
\affiliations{
    \textsuperscript{\rm 1} University of Florida, \\
    \textsuperscript{\rm 2} Carnegie Mellon University \\

    mkunzlermaldaner@ufl.edu,
    hanwend@cs.cmu.edu
}

\begin{document}

\maketitle

\begin{abstract}
While generative AI systems have gained popularity in diverse applications, their potential to produce harmful outputs limits their trustworthiness and usability in different applications. Recent years have seen growing interest in engaging diverse AI users in auditing generative AI that might impact their lives. To this end, we propose MIRAGE as a web-based tool where AI users can compare outputs from multiple AI text-to-image (T2I) models by auditing AI-generated images, and report their findings in a structured way. We used MIRAGE to conduct a preliminary user study with five participants and found that MIRAGE users could leverage their own lived experiences and identities to surface previously unnoticed details around harmful biases when reviewing multiple T2I models' outputs compared to reviewing only one. 
\looseness=-1
\end{abstract}

\begin{figure}[ht]
  \centering
  \includegraphics[width=\columnwidth]{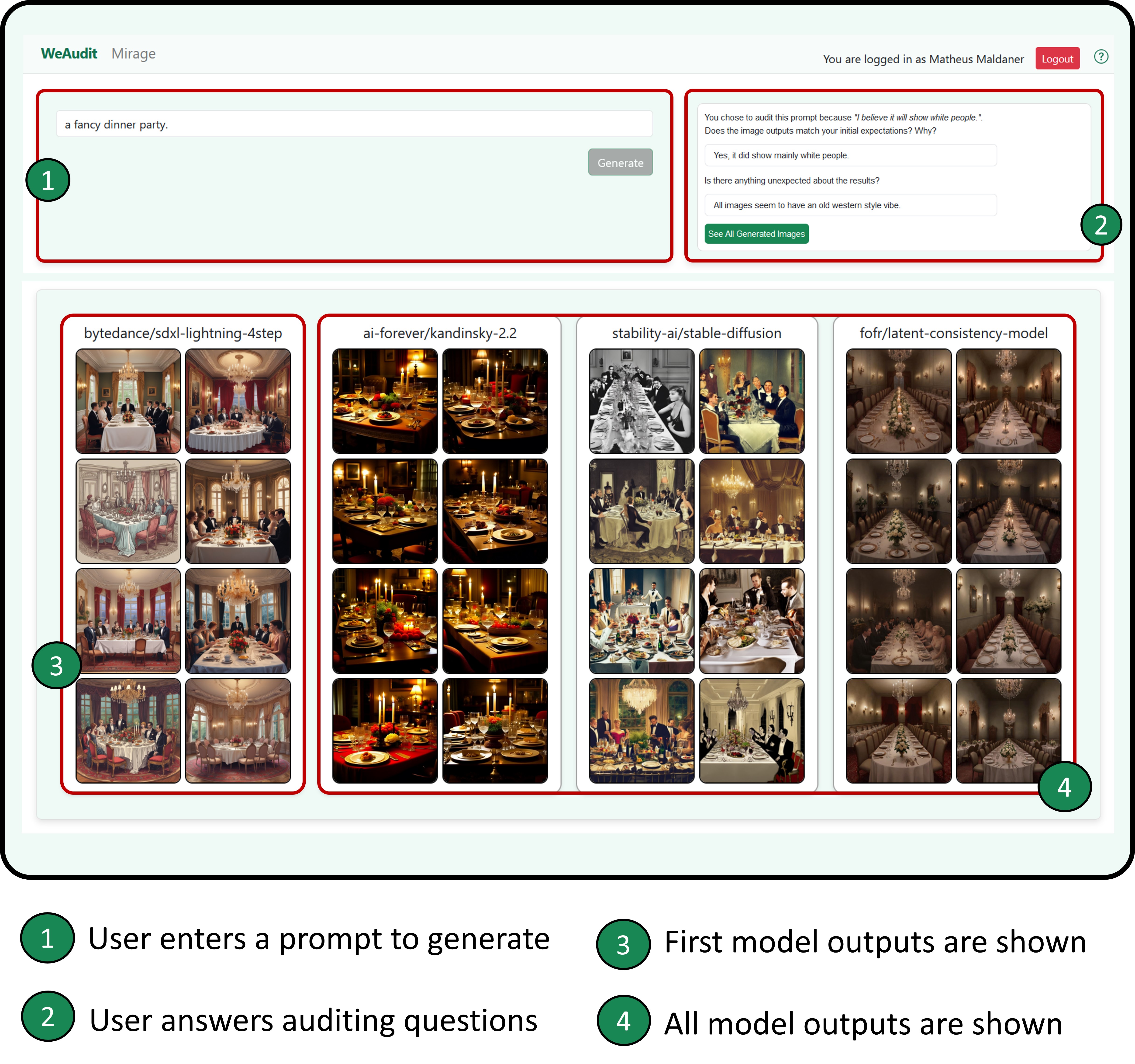}
  \caption{MIRAGE User Interface and Outlined Workflow}
  \label{fig:teasing}
\end{figure}

\section{Introduction}
Despite their popularity, generative AI such as text-to-image (T2I) systems can lead to problematic outputs, such as harmful biases reinforcing societal stereotypes or producing misleading information \cite{bianchi2023easilyaccessiblet2i}. For example, a recent news article suggested that Stable Diffusion, a popular open-source T2I model, rarely depicted women as doctors, lawyers, or judges, and often suggested that men with dark skin commit crimes, reinforcing harmful gender and racial biases \cite{Nicoletti2023humans, buolamwini2018gender}.

Auditing algorithms are crucial for detecting discrimination and ensuring fairness in AI systems \cite{Sandvig2014AuditingA}.
Recognizing the power of diverse end users in surfacing harmful behaviors in AI systems that might otherwise be overlooked by small groups of AI developers, recent research has explored engaging users in auditing AI systems \cite{shen2021everydayalgorithmauditing, devos2022toward}. Industry AI teams are also often motivated to engage diverse users in auditing their AI systems and products, and often employ crowdsourcing platform to recruit and assign auditing tasks to users \cite{deng2023understanding}. However, they often lack effective mechanisms to scaffold users in surfacing harmful biases that are relevant to their own identities and lived experiences \cite{Lam2022endusersaudits}.

In this paper, we explore whether \textbf{viewing outputs from multiple text-to-image models} can help users identify image details they might have missed in a single model output. To explore this, we contribute (1) the development of MIRAGE, a web-based interface for multi-model side-by-side comparison, (2) preliminary findings from five user studies, and (3) the exploration of future work directions for MIRAGE and everyday user auditing \cite{shen2021everydayalgorithmauditing}.

\section{Related Work}

A small but growing line of work has explored developing tools for engaging AI developers and users in testing and auditing AI models. For example, Lam et al. developed IndieLabel, an end-user audit system, to empower users with low tech-savviness in auditing sentiment analysis algorithms \cite{Lam2022endusersaudits}. Kahng et al. designed LLM comparators for AI developers to more effectively compare and audit text-based LLM outputs \cite{kahng2024llmcomparatorvisualanalytics}.

More recently, Replicate, a platform providing an array of open-source generative AI models, has developed Zoo as an open-source text-to-image playground \cite{Replicate, Zoo}. Although Zoo allows users to compare different model outputs, the interface is not particularly designed with auditing as its primary goal \cite{Zoo}. For example, users are unable to provide feedback on the model outputs. In addition, Zoo’s model outputs are laid out vertically, requiring users to scroll to see different results, which reduces efficiency in directly comparing model outputs \cite{tufte2001visual}.

Our work extends this prior work by designing, developing, and evaluating a web-based interface that allows AI users to review multiple T2I model outputs and report their audit insights in a structured way.

\section{System Description}

MIRAGE is a web-based interface designed to facilitate the auditing of generative text-to-image AI models. Upon entering MIRAGE, users first input a prompt they would like to audit (Figure \ref{fig:teasing}, Step 1). They will then see image outputs from the bytedance/sdxl-lightning-4step model, chosen for its fast inference speed and high image quality which ensures minimal wait times (Figure \ref{fig:teasing}, Step 3). After auditing the single model's image outputs, participants proceed to audit multiple image outputs from all four predefined models: bytedance/sdxl-lightning-4step, ai-forever/kandinsky-2.2, stability-ai/stable-diffusion, and fofr/latent-consistency-model (Figure \ref{fig:teasing}, Step 4). Between these steps, participants answer prepared questions shown in Figure \ref{fig:workflow} that will ultimately become part of the audit report. Please visit \url{mirage.weaudit.org} to access MIRAGE \footnote{We plan to live demo MIRAGE during the conference}.

\begin{figure}[t]
\centering
\includegraphics[width=0.9\columnwidth]{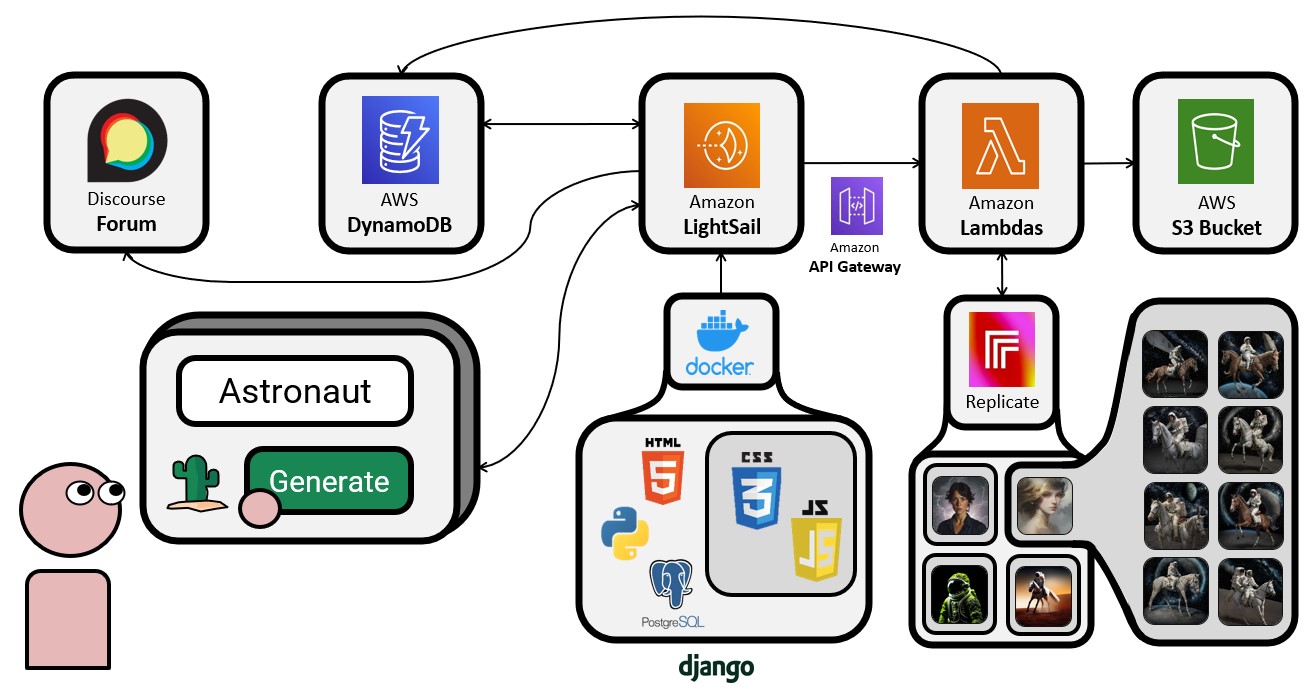}
\caption{MIRAGE Technical Implementation}
\label{fig:backend}
\end{figure}

\section{Implementation}

In order to achieve our goal, we first developed the MIRAGE web application (Figure \ref{fig:backend}), hosting it on Amazon Lightsail with a Docker container packaging a Django project. Image generation for specific models is handled through API calls to Replicate, a centralized service that works as a hub for open-source models. After generation, images are stored in an Amazon S3 Bucket with unique IDs and referenced in an Amazon DynamoDB table for easy retrieval. Backend computations are performed using AWS Lambda functions, accessed through Amazon API Gateway. Although not used in our study, we also integrate a discussion forum developed using the Discourse API, so that users can choose to post and discuss their audit findings.

\section{Preliminary User Study}

To test our hypothesis that multi-model comparisons can help users detect biases, we conducted an initial user study with five participants to evaluate the effectiveness of MIRAGE. Participants were first asked to enter a predefined prompt (``a fancy dinner party'') but later had the opportunity to explore the tool by entering two prompts of their choice, where we encouraged them to consider their background and identities. While the models were generating images in the background, participants were asked additional questions to reflect on their lived experiences and the images they anticipated. 

\begin{figure}[t]
\centering
\includegraphics[width=1\columnwidth]{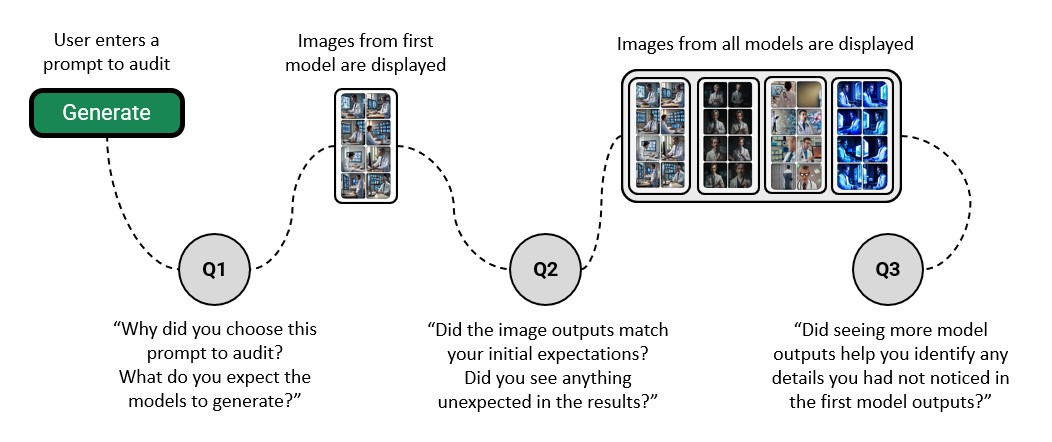} 
\caption{MIRAGE User Study Workflow}
\label{fig:workflow}
\end{figure}

As shown in Figure \ref{fig:workflow}, participants were first asked why they chose the specific prompt to audit and what they expected the models to generate. This step encouraged users to reflect on their own biases while allowing all models to generate outputs in the background without the users noticing. Initially, participants were only shown the image outputs of the first model and asked whether these matched their initial expectations and if there was anything unexpected about the results. After answering these questions, users were shown all four model outputs, given time to examine them, and then asked whether seeing more models helped them identify any potentially harmful details exhibited in the first model's outputs that they had previously missed.

For this preliminary user study, we recruited our participants through convenient sampling. The study was approved by our Institutional Review Board. Each session lasted around 30-45 minutes, and all participants were compensated with \$15 Amazon gift cards for their time.

\section{Preliminary Findings}

The preliminary user study of MIRAGE provided several insights into its usefulness and usability. To start with, all five participants found MIRAGE easy to use, with a clear and intuitive workflow. P3 especially appreciated the question flow (see Figure \ref{fig:workflow}), which allowed them to reflect on their own biases and articulate the reasons behind their prompt choices for auditing.

In addition, viewing outputs from additional models enabled participants to identify new image details that could lead to potentially harmful biases, which were not recognized when viewing the outputs of a single model. For instance, when P1 saw the first model's output for the prompt ``An intern working in the city of Pittsburgh,'' they only noted that all workers were depicted outdoors. However, after seeing additional model outputs, they observed that darker-skinned workers were depicted wearing construction clothes, while lighter-skinned workers were depicted in office attire, carrying keyboards and notebooks. In another example, participants who initially expected the prompt ``A fancy dinner party'' to generate images of white males noticed that all tables were long and rectangular, a more Western style, as opposed to the round tables commonly found in Asian cultures.

The exposure to multiple images led participants to develop new auditing strategies. In particular, P1, P4, and P5 reported that they focused on inspecting individual images when there was only one model but focused on reviewing the overall image output distribution when there were multiple models.

Finally, participants felt that reviewing multiple models allowed them to more effectively assess output styles and model quality, shedding light on new future directions for the tool which are explored below.

\section{Future Use Cases}

While our user study using MIRAGE focused on spotting biases and harmful behaviors in T2I systems, our preliminary findings suggest several future work directions, which we elaborate on below.

\textbf{(1) Anonymous Auditing}; Companies and developers are often reluctant to give public access to their AI models due to the risk of exposing proprietary technology to competitors or facing public backlash if the models produce harmful content \cite{deng2023understanding}. This reluctance limits the ability of companies to gather valuable feedback from everyday users, who are social actors engaged in the daily use of algorithm auditing systems \cite{shen2021everydayalgorithmauditing}. In line with a functionality provided by Chatbot Arena \cite{chiang2024chatbotarenaopenplatform}, we envision adapting MIRAGE to act as a bridge between everyday users and developers of proprietary text-to-image models. In this future application, developers can submit their models to MIRAGE, which will anonymize and deliver them to everyday users who can leverage their lived experiences and social backgrounds to provide feedback and locate harmful behaviors. 

\textbf{(2) Text-to-Image Model Supermarket}; Preliminary findings from our user studies found that participants naturally started to draw conclusions about the characteristics and overall style of each model. This opens the possibility for a ``text-to-image model supermarket'' where users can enter a prompt and quickly visualize outputs from many text-to-image models side-by-side. 
We envision a future version of MIRAGE that could enable users to rapidly understand the capabilities and limitations of different T2I systems.
Users can then choose different models depending on their use case. For example, a parent creating a bedtime story for their children might prefer cartoonish images, while a businessman might want more realistic images.

\textbf{(3) Text-to-Image Model Leaderboard}; We seek to explore a model leaderboard system similar to Chatbot Arena \cite{chiang2024chatbotarenaopenplatform} but focused on T2I models. This system would allow users to enter their prompt and select which output best aligns with their expectations, working as a way to rank models and encourage developers to take user feedback into account when developing or fine-tuning models. A leaderboard system would promote healthy competition among model developers, encouraging them to improve their models based on real user feedback. This could lead to rapid advancements in model quality and ethical considerations, as developers seek to address biases and other issues highlighted by users. A leaderboard can also be easily incorporated into the existing AI practitioners’ workshop, which prior research has repeatedly identified as an important factor for Responsible AI toolkits to be adopted by AI teams in practice \cite{deng2022exploring, yildirim2023investigating}. Additionally, researchers could use the data collected from this leaderboard system to study user preferences and gain insights into common biases and areas for improvement in T2I models. 

\section{Conclusion}
In this paper, we introduced MIRAGE, a web-based tool designed to facilitate side-by-side comparisons of multiple AI text-to-image models. Our preliminary user study, involving five participants, showed that MIRAGE was successful in helping users identify biases and harmful behaviors in model outputs. Participants appreciated the clear and intuitive workflow of MIRAGE and were able to discover new details and potential biases that were not apparent when viewing the outputs of a single model. Looking ahead, we propose several future use cases for MIRAGE, including the development of anonymous auditing, the creation of a text-to-image model supermarket, and the establishment of a text-to-image model leaderboard system. These directions aim to bridge the gap between everyday users and developers and create a more inclusive and effective AI model auditing ecosystem.

\section{Acknowledgments}

\noindent We would like to thank Claire Wang, Serena Cai, Ashish Shugani, and Bhavya Jha for their continuous support in the development of MIRAGE. We also extend our gratitude to Raul Valle, Autumn Qiu, and Evan Partidas for their help reviewing the paper. This work was supported by the National Science Foundation (NSF) program on Fairness in AI in collaboration with Amazon under Award No. IIS-2040942.

\clearpage

\bibliography{citation}

\end{document}